\documentclass[twocolumn]{aastex61}

\newcommand{\dom}[1]{{#1}} 
\submitjournal{ApJL}

\shorttitle{Reionization of z=0 halos in Coda I-AMR}
\shortauthors{Aubert et al.}

\begin{document}

\title{The Inhomogeneous Reionization Times of Present-day Galaxies}

\correspondingauthor{Dominique Aubert}
\email{dominique.aubert@astro.unistra.fr}

\author{Dominique Aubert}
\affil{Universite de Strasbourg, CNRS, Observatoire astronomique de Strasbourg, UMR 7550, F-67000 Strasbourg, France}

\author{Nicolas Deparis}
\affil{Universite de Strasbourg, CNRS, Observatoire astronomique de Strasbourg, UMR 7550, F-67000 Strasbourg, France}

\author{Pierre Ocvirk}
\affil{Universite de Strasbourg, CNRS, Observatoire astronomique de Strasbourg, UMR 7550, F-67000 Strasbourg, France}

\author{Paul R. Shapiro}
\affil{Department of Astronomy, University Texas, Austin, TX 78712-1083, USA}

\author{Ilian T. Iliev}
\affil{Astronomy Center, Department of Physics \& Astronomy, Pevensey II Building, University of Sussex, Falmer, Brighton BN1 9QH, United Kingdom}

\author{Gustavo Yepes}
\affil{Departamento de Fisica Teorica and CIAFF, Modulo M-15, Universidad Autonoma de Madrid, Cantoblanco 28049, Spain.}

\author{Stefan Gottl\"{o}ber}
\affil{Leibniz-Institute f\"{u}r Astrophysik Potsdam (AIP), An der Sternwarte 16, D-14482 Potsdam, Germany}

\author{Yehuda Hoffman}
\affil{Racah Institute of Physics, Hebrew University, Jerusalem 91904, Israel}

\author{Romain Teyssier}
\affil{Institute for Theoretical Physics, University of Zurich, Winterthurerstrasse 190, CH-8057 Z\"urich, Switzerland}

\begin{abstract}
Today's galaxies experienced cosmic reionization at different times in different locations. For the first time, reionization ($50\%$ ionized) redshifts, $z_R$, at the location of their progenitors are derived  from new, fully-coupled radiation-hydrodynamics simulation of galaxy formation and reionization at $z > 6$, matched to N-body simulation to z = 0. Constrained initial conditions were chosen to form the well-known structures of the local universe, including the Local Group and Virgo,
in a (91 Mpc)$^3$ volume large enough to model both global and local reionization. Reionization simulation CoDa I-AMR, by CPU-GPU code EMMA, used (2048)$^3$ particles and (2048)$^3$ initial cells, adaptively-refined, while N-body simulation CoDa I-DM2048, by Gadget2, used (2048)$^3$ particles, to find reionization times for all galaxies at z = 0 with masses $M(z=0)\ge 10^8 M_\odot$. 
Galaxies with $M(z=0) \gtrsim 10^{11} M_\odot$ reionized earlier than the universe as a whole, by up to $\sim$ 500 Myrs, with significant scatter. For Milky-Way-like galaxies, $z_R$ ranged from 8 to 15. Galaxies with $M(z=0) \lesssim 10^{11} M_\odot$ typically reionized as late or later than globally-averaged $50\%$ reionization at $\langle z_R\rangle =7.8$, in neighborhoods where reionization was completed by external radiation. The spread of reionization times within galaxies was sometimes as large as the galaxy-to-galaxy scatter. The Milky Way and M31 reionized earlier than global reionization but later than typical for their mass, neither dominated by external radiation.
Their most massive progenitors at $z>6$ had $z_R$ = 9.8 (MW) and 11 (M31), while their total masses had $z_R$ = 8.2 (both).
\end{abstract}




\section{Introduction}


Different patches of the universe reionized at different times, over a wide range of redshifts and this local reionization time left its imprint on galaxies at $z = 0$. Reionization photoheating suppressed baryonic infall and star formation in low-mass 
galaxies and caused reionization to self-regulate (e.g. \citet{SGB94}, \citet{ILIEV07}).
The stellar populations of their satellites, for example, were dramatically affected by when reionization occurred and whether instantaneous or extended (see e.g.
\citet{KOP9,BUS10,OCV11,OCV14}).
Reionization suppression is thought to reconcile the observed paucity of satellites in the Local Group (LG) with their over-prediction by N-body simulations of $\Lambda$CDM \citep{BUL17}. In a global context, where the contribution of low-mass galaxies to reionization is still debated (see e.g. \citet{BOU14,FIN15}), these effects must be understood in order to interpret observations of high-z galaxies. 

{For the first time, we are able to perform a complete study of the reionization history of z=0 galaxies and the LG in particular, using a full-physics simulation. 
\dom{We produced a new, state-of-the-art, fully-coupled radiative hydrodynamics (RHD) simulation of galaxy formation and reionization at $z > 6$, named CoDa I-AMR ("Cosmic Dawn"), by CPU-GPU {adaptive mesh refinement (AMR)} code EMMA \citep{AUB15}. {This new code is able to take advantage of the GPU-driven hybrid architecture of Titan supercomputer (ORNL) and required 20 million core hours to produce this full physics RHD simulation.  We combined CoDa I-AMR with a dark-matter-only N-body simulation to $z = 0$ by Gadget2, 
CoDa I-DM2048, from the same initial conditions (ICs), to match today's galaxies with their reionization histories self-consistently. Reionization simulation CoDa I-AMR used (2048)$^3$ particles and (2048)$^3$ initial cells, while N-body simulation CoDa I-DM2048, used (2048)$^3$ particles,  to find reionization times and durations for all the galaxies in a {(64 $h^{-1}$ Mpc $\sim$ 91 Mpc)$^3$ volume at z = 0 with  $M\ge 10^8 h^{-1} M_\odot$.}}}}

{These ICs are a constrained realization of $\Lambda$CDM, constructed by the CLUEs ("Constrained Local UniversE Simulations") project from observations of galaxies in the local universe, to form familiar structures within it, including the LG with the Milky Way (MW) and M31, in a volume large enough to model both global and local reionization. 
 The reionization history of the LG in its authentic environment is thereby modelled, to assess how representative it is, as the most accessible place to observe galaxies and their satellites today to deduce their histories.  We are therefore able to compare the LG to a population of analog galaxies in different environments and to establish, as shown hereafter, that LG galaxies are reionized later than reference galaxies of similar masses and that the LG is reionized without influence from external incoming fronts. }

{\citet{OCV16} reported our first attempt to model both
global reionization and its impact on the LG, using the first RHD simulation of reionization of the Local Universe (CoDa I) with high-enough resolution in a large-enough volume. \dom{This first breakthrough used CPU-GPU code Ramses-Cudaton with $(4096)^3$ particles and $(4096)^3$ cells on a unigrid, in a (91 Mpc)$^3$ volume, from the same ICs used here.  However it used a subgrid model of star formation for which the choice of efficiency parameters made reionization finish later ($z\sim  4.5$) than observations of global reionization suggest, thereby making it unsuitable for direct comparison with the Local Universe today. With CoDa I-AMR, we report here a new breakthrough. Our new CPU-GPU code EMMA uses AMR methodology to further increase force resolution where required and beyond that of the unigrid used in \citet{OCV16}. It also uses an improved calibration of subgrid stellar physics model to finish reionization by z = 6, as required by observation of global reionization \citep{FAN6}.}} 

{We also overcome key limitations of our previous studies on the reionization of z=0 galaxies. \citet{OCV14} connected reionization histories with z=0 LG galaxies using zoom simulations focused on (Mpc)$^3$ volumes and post-processing radiative transfer (PPRT). These small volumes cannot account for the influence of distant powerful emitters
 and PPRT cannot model e.g. the radiative suppression of star formation. \citet{WEI07} and  \citet{DIX17}  also modeled the large volumes with  PPRT, at lower spatial resolution ($256^3$ vs $2048^3$ + AMR here) and use pre-defined halo emissivities, impacting the small-scale and stochastic details of galaxies reionizations (reionization durations, scatter of reionization times). Our study also complements \citet{ALV9} and \citet{LI14}, who focused on scales relevant to massive galaxies and clusters, using semi-numerical methodology. The scales explored here allows to focus on smaller masses between $10^8 h^{-1} M_\odot$ and $10^{13} h^{-1} M_\odot$. }

\dom{In summary, we study in this letter the reionization history of z=0 galaxies. For this purpose, we used the new CPU-GPU AMR code EMMA for the first time to produce a self-consistent and fully-coupled RHD simulation of galaxy formation and reionization, from constrained-realization ICs of the Local Universe. The (91 Mpc)$^3$ volume is large enough to model global reionization, and the  resolution is sufficient to study all galaxies with $M \ge 10^8 h^{-1} M_\odot$.   A companion N-body simulation from the same ICs is then used to project forward in time, allowing us to identify all the z = 0 galaxies in that volume today, including the MW and M31. The LG, and its reionization history, can therefore be compared to the population of galaxies, within a single consistent framework. We describe our simulations and their analysis in Section 2. Our results for the reionization times and durations of these simulated z = 0 galaxies are presented for the general population in Sections 3.1 and 3.2, before discussing the LG in Section 3.3.}





\section{Methodology}

\subsection{Initial Conditions}

The CLUES constrained-realization ICs used here assume a WMAP 5 cosmology ($\Omega_m=0.279, \Omega_v=0.721, H_0=70$ km/s/Mpc, \citet{HIN9}) in a (64$h^{-1}$ Mpc $\sim$ 91 Mpc)$^3$
comoving volume with $2048^3$ particles and cells. These ICs are a coarsened version of those in CoDa I \citep{OCV16}. Initial phases were chosen to reproduce the observed structures of the local universe at z=0, providing an MW - M31 pair with the right mass range and separation in the proper large-scale environment (see \citet{GOT10}, \citet{ILI11}, \citet{FR11}). 



\subsection{Reionization Simulation to $z = 6$}
EOR simulation CoDa I-AMR to $z = 6$, by hybrid CPU-GPU, AMR code EMMA \citep{AUB15}, solved fully-coupled equations of collisionless DM dynamics, hydrodynamics and radiative transfer, with standard sub-grid models for star formation and supernova feedback (\citet{RAS06}, \citet{DEP17}). 
The initial spatial grid of $2048^3$ cells was refined whenever a cell contained more than 8 DM particles, until cell-widths reached 500 pc (proper), corresponding to three refinement levels by z=6, increasing total cell number by 2.1, to 18 billion.   CoDa I-AMR was produced on Titan (ORNL) using 32768 CPU cores and 4096 GPUs dedicated to RHD solvers.

Star formation was triggered if gas overdensity in a cell exceeded 50, resulting in first star particles at $z\sim 19$; once triggered, star formation obeyed a Schmidt-Kenicutt Law with efficiency $1\%$ (see \citet{RAS06}, \citet{DEP17}). Star-particles, of mass  $7\times 10^4 M_\odot$, released ionizing photons according to a Starburst 99 population model \citep{LEI99} with Top-Heavy IMF and 0.05 $Z_\odot$ metallicity; the corresponding emissivity was $1.5\times 10^{17}$ ionising photons/sec/stellar kg for $3\times 10^6$ years, followed by exponential decline. A sugrid escape fraction $0.2$ was applied to compute the photon number released inside the star-particle's cell.  
This ensured that global reionization finished at $z\sim 6$.  

{Radiative transfer was moment-based, with M1 closure and a reduced speed of light $c_\mathrm{sim}=0.1$}. Mechanical feedback from supernovae was included, with energy $9.8\times 10^{11}$ J/stellar kg injected into
surrounding gas after $15\times 10^6$ years: 1/3 via thermal energy, 2/3 via kinetic winds. At z=6, $120\times 10^6$ star-particles were present.



\begin{figure}[ht!]
\includegraphics[height=1.3\columnwidth, width=1.\columnwidth]{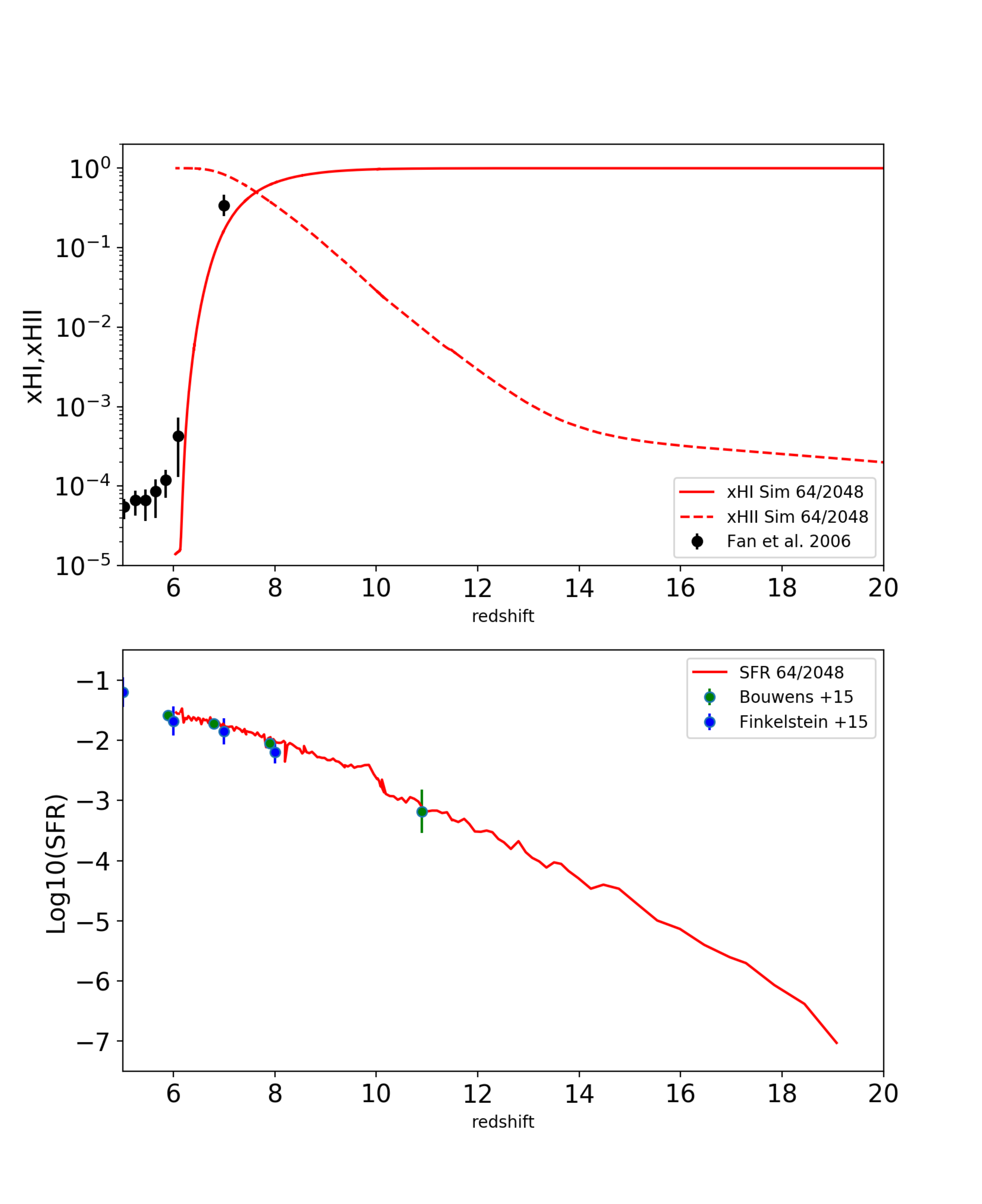}
\caption{Globally-averaged results of CoDa I-AMR simulation vs. redshift:  volume-averaged ionized/neutral fraction (top), cosmic star formation rate ($M_\odot/\mathrm{yr}/\mathrm{Mpc}^3$, bottom).}
\label{fig:globpro}
\end{figure}

Global reionization in CODA I-AMR finished by $z=6.1$, {with a Thompson optical depth $\tau=0.069$ (assuming pure hydrogen) consistent with} (\citet{PLA15}, $0.066 \pm 0.016$) but with a residual neutral fraction lower than expected from quasar data (\citet{FAN6}, see Fig. \ref{fig:globpro}). The average star formation history was consistent with that inferred from observations of the UV luminosity function of high-z galaxies \citep{BOU14,FIN15}.  

\subsection{Dark-Matter-Only Simulation to $z = 0$}

The properties of z=0 halos were obtained from a dark-matter-only N-body simulation, CoDa I-DM2048, by Gadget2 \citep{SPR5}, from the same ICs as CoDa I-AMR. 
Halos were identified using a FOF algorithm with linking length 0.2 and a minimum number of particles of 10, leading to $\sim 20$ million halos identified at z=0. The smallest-mass FOF objects detected had $2.4\times 10^7 M_\odot$. Merger trees were also generated to connect z=0 halos to their progenitors during the EoR \citep{RIE13}. 



\subsection{Reionization maps}
\begin{figure*}[ht]
\begin{center}
\includegraphics[width=2.3 \columnwidth]{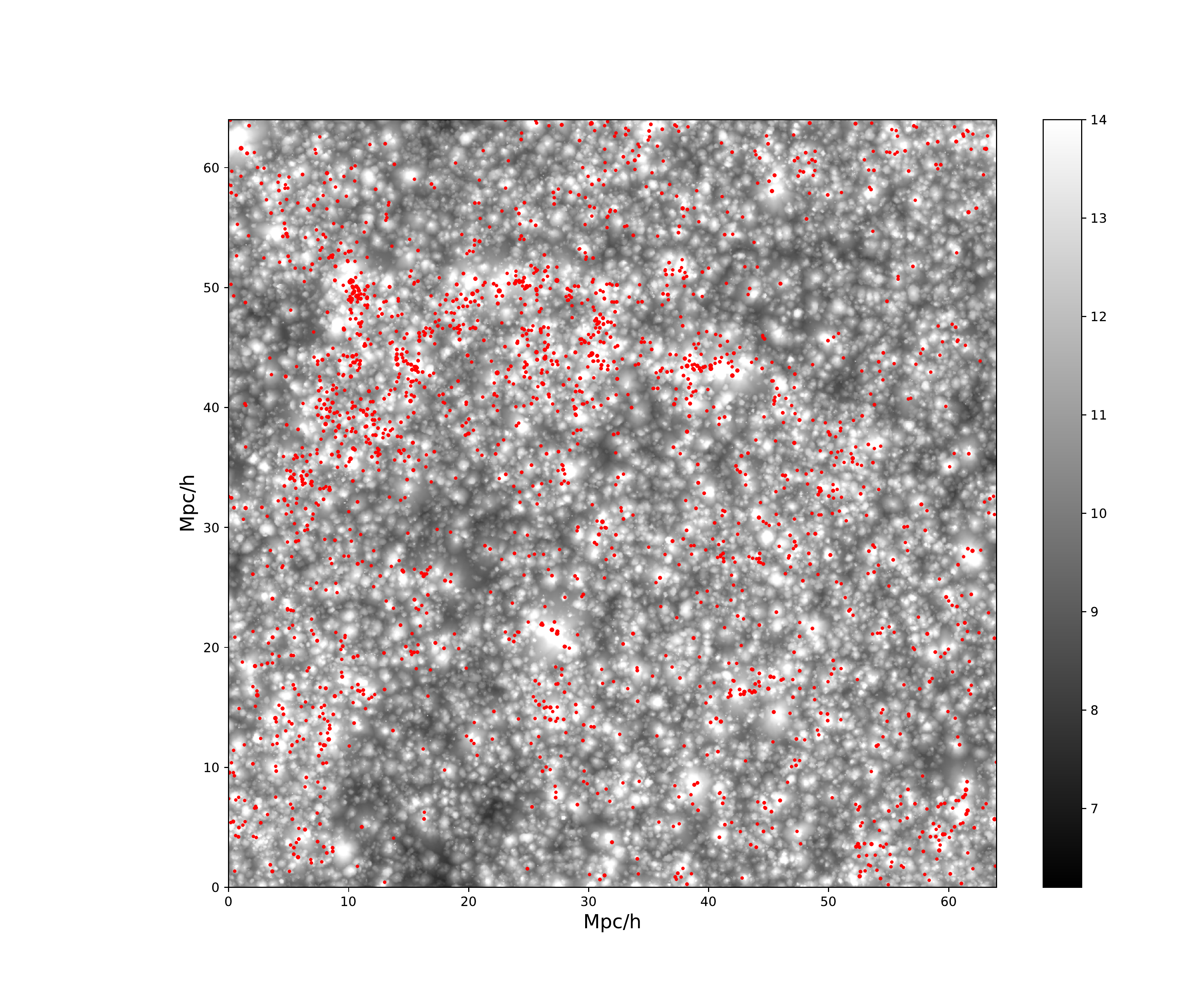}
\caption{Projected distributions of the 2000 most massive halos in DM-only simulation CoDa I-DM2048 at z=6 (symbols) and the maximum values of the reionization redshift from CoDa I-AMR, along the projection axis.}
\end{center}
\label{fig:reion_halo_map}
\end{figure*}

A 3-D map of reionization times (and redshifts) was created from the evolving, inhomogeneous, nonequilibrium ionization state of hydrogen in CoDa I-AMR.  We define the reionization time as the instant when a cell first crossed the ionized-fraction threshold 0.5. The time resolution was 1.4 Myrs at $z = 6$.   
The result is a 3-D field, $t_\mathrm{reion}(x,y,z)$, sampled using $2048^3$ pixels corresponding to the base resolution of our simulation (see Fig. \ref{fig:reion_halo_map}). 
There is a clear correlation between the CoDa I-DM2048 halo distribution at z=6 and the CODA I-AMR reionization map : {halos are found at the centers of ionized patches and their spatial distribution matches the topology of $t_\mathrm{reion}(x,y,z)$}.


Cell-based reionization times do not distinguish cells inside galactic halos from intergalactic cells.  It is the IGM, however, which undergoes reionization, while interstellar gas inside galaxies may remain neutral even after global reionization ends.  It is, therefore, the reionization time of the IGM at the locations of the progenitor halos or particles we seek, specifically the times at which these locations first reached ionized fraction 0.5.  Henceforth, this is what we shall mean when we assign a 'galaxy reionization time' $t_\mathrm{reion}$ to a $z = 0$ galaxy.


\subsection{Progenitor-based Reionization Times }
We start by assigning reionization times to $z = 0$ galaxies using those of their progenitor halos, using the merger-trees of the DM-only simulation to determine the positions, $ \vec x_\mathrm{mm}$, at each $z > 6$, of their most-massive (mm) progenitors. Starting from $z = 20$, we find the earliest step in the merger tree when the most massive progenitor at that step belongs to an ionized cell. This step has redshift $z_R$, and the reionization time of a $z = 0$ halo is given by
\begin{equation}
t_\mathrm{prog}=t_\mathrm{reion}(\vec x_\mathrm{mmR}),
\end{equation}
where $\vec x_\mathrm{mmR}$ is the center-of-mass position of the most massive progenitor of this halo at $z=z_R$.

A halo is assigned a reionization time only if it has a progenitor at $z > 6$; this is not the case if progenitor halos only emerged after $z = 6$ or were not detected by the FOF algorithm before this.  
However, this procedure guarantees that $t_\mathrm{prog}$ is set by material already in the structure by $z > 6$. {Though they only represent $5\%$ of $z=0$ halos, all halos with $M_{z=0} > 5\times 10^9 h^{-1} M_\odot$ have progenitor-based reionization times.}


\subsection{Particle-based Reionization Times}
Our second approach is based upon the reionization times of all the DM particles that belong to a halo at $z = 0$.
 Once these particles are identified, their positions at $z > 6$ can be traced using simulation snaphots. Each particle is then assigned a reionization time defined as the earliest time $t(z_R)$ it was located inside an ionized cell in the reionization map. An average particle-based $\langle t_\mathrm{part} \rangle$ 
 is then assigned to each halo:
\begin{equation}
\langle t_\mathrm{part}\rangle=\frac{\sum_{\vec x_\mathrm{p0} \in \mathrm{halo} \ } t_\mathrm{reion}(\vec x_\mathrm{pR})}{\sum_{\vec x_\mathrm{p0} \in \mathrm{halo} \ } 1},
\end{equation}
where $\vec x_\mathrm{p0}$ and $\vec x_\mathrm{pR}$ are 
particle positions at $z = 0$ and $z = z_R$, respectively.

This procedure is more difficult, as it requires cross-matching $8\times 10^9$ DM particles with $\sim 20\times 10^6$ $z = 0$ halos to identify the particles belonging to each halo. 
However, this technique has the advantage that it assigns reionization times to all $z = 0$ halos, even the smallest. Reionization times determined this way tend to be later than those by the other method since diffuse material, presumably reionized at later times and/or accreted after reionization, is included.
{We also computed for each halo the time at which 50\%, 10\%, 1\% and 0.1\% of its particles have been reionized. } 
%
\section{Results}

%
%
%
%

\subsection{Reionization Times}

\begin{figure}[ht]
\includegraphics[height=1.\columnwidth,width=1.1\columnwidth]{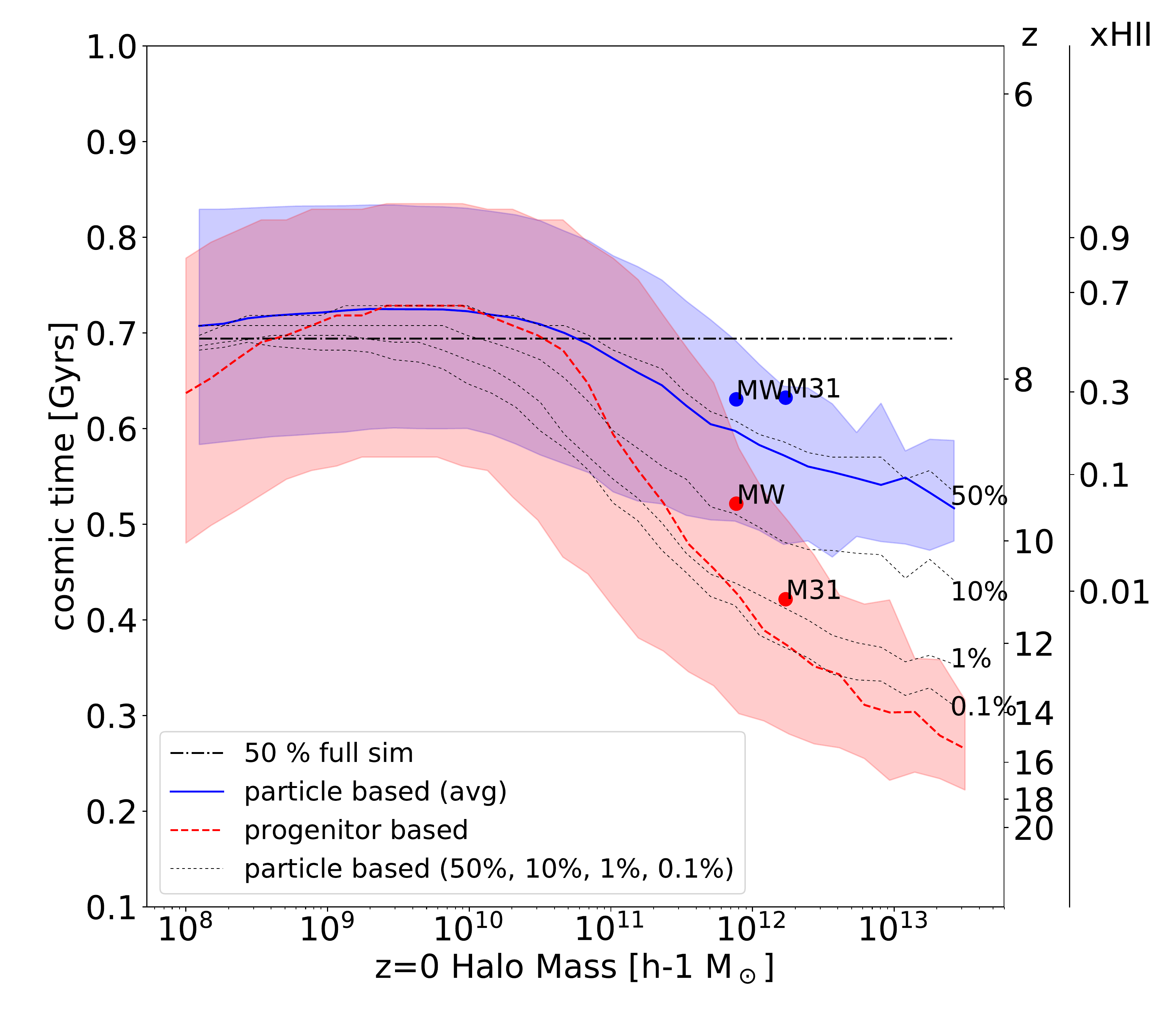}
\includegraphics[width=1.1\columnwidth]{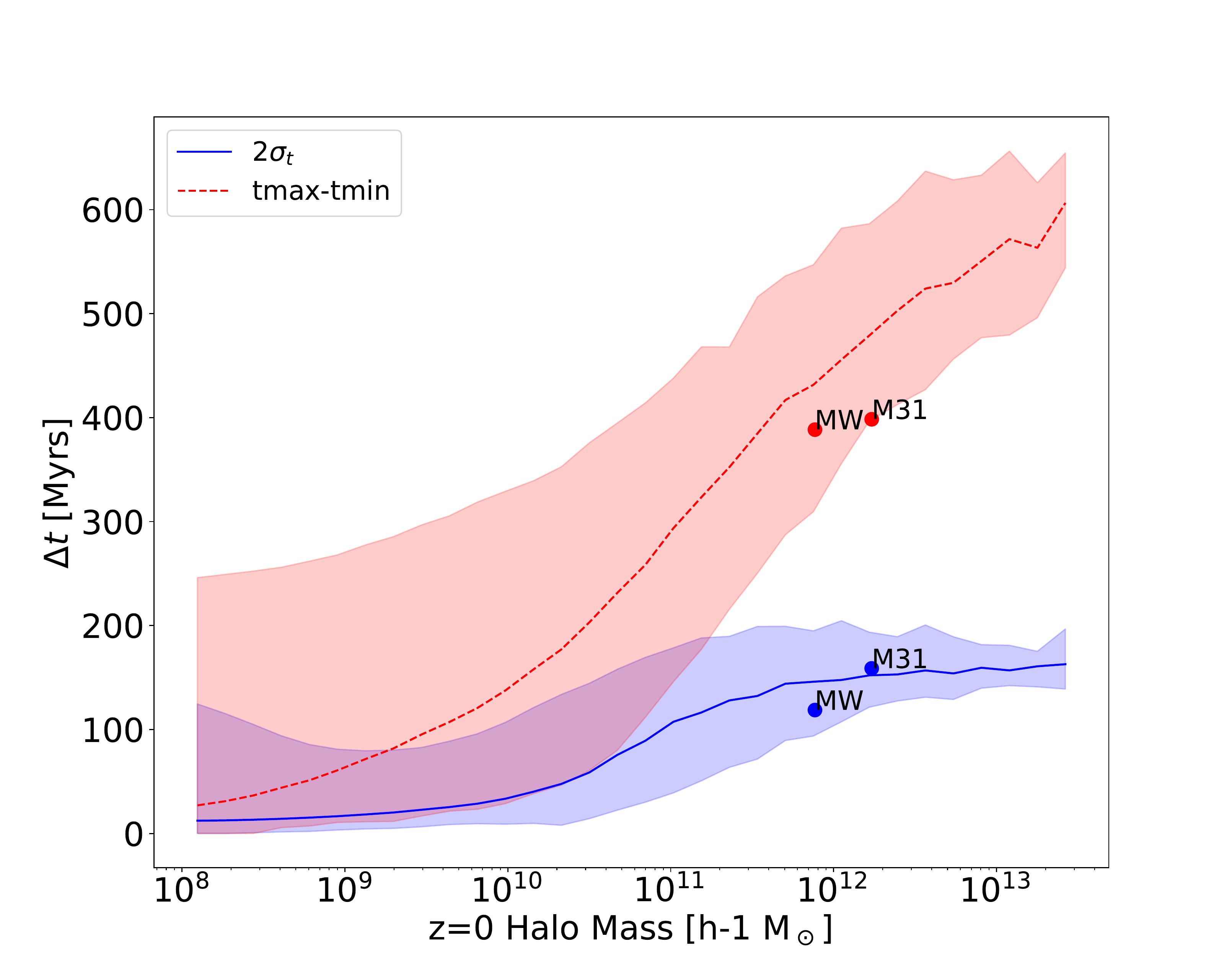}
\caption{\textit{Top}: Reionization times vs. present-day halo mass.
Full volume reaches $x_{HI}=0.5$ at $z\sim 7.8$ (dashed line).
 \textit{Bottom}: Reionization durations $\Delta t$ vs. present-day halo mass. In both panels, lines stand for the median value within each mass bin, and shaded areas span the $5\%-95\%$ percentiles. Dots indicate the values for the simulated MW-M31 pair. See text regarding the different types of measurements.}
\label{fig:treion}
\end{figure}

Reionization times of z=0 halos are shown in Fig. \ref{fig:treion}. Galaxies with $M_{z=0}>10^{11} h^{-1}M_\odot$ reionized up to $500$ Myrs earlier than the full volume, depending on the method used. In this mass range, the more massive the galaxies, the earlier they reionized, as expected since their progenitors were located in denser environments where more intense sources formed. For lower masses, ($M_{z=0}<10^{11} h^{-1}M_\odot$), reionization times were typically consistent with the global one.  Their median time was slightly later than the global time : these objects were fainter or even star-less and were externally reionized. Since their immediate environment was denser than the average IGM, they ionized later than the IGM.  Nevertheless, halo-to-halo scatter is significant ($\sim 250$ Myrs $5\%-95\%$ percentile).

Results for the two methods are consistent but different. The progenitor-based technique applies only to objects already formed at high z : it finds the reionization redshift of the oldest material of a z=0 halo, thus explaining why it consistently yields lower $t_\mathrm{reion}$. On the other hand, it requires that FOF objects pre-existed at $z>6$, biasing the halo sample:  $10^8 h^{-1} M_\odot$ halos at z=0  must have had peculiar accretion rates to have progenitors at $z>6$ and low mass at z=0. The dip in reionization times at the low-mass end confirms this, and our results indicate these objects are located in high-density regions, thus explaining their low $t_\mathrm{reion}$. It may also indicate these objects were more massive in the past and were stripped : these low-mass objects were assigned $t_\mathrm{reion}$ typical of more massive objects.

The particle-based method suffers less from this bias because all z=0 halos are included~: the $t_\mathrm{reion}$ dip at the low-mass end disappears. It returns lower reionization redshifts for $M_{z=0}>10^{11} h^{-1} M_\odot$, resulting from the fraction of material in halos at z=0 that was diffuse matter in the IGM and reionized at later times. {The times at which 50\%, 10\%, 1\% and 0.1\% of particles were reionized show a hierarchy : 50\% reionization times are consistent with average particle-based values while 0.1\% values provide earlier $z_r$. Median progenitor-based reionization times are recovered assuming a smaller percentile of particles for larger $M_{z=0}$, corresponding to a typical 'reionization mass' $\sim 10^{10} h^{-1}M_\odot$. }

\citet{LI14} found that $10^{12} h^{-1} M_\odot$ galaxies tend to reionize earlier than the IGM, by $\Delta z\sim 1\pm1$, while we find earlier reionization times for these objects, too, but by $\Delta z\sim 1.5\pm 1.5$ (particle-based) or $\Delta z\sim 4.5\pm 3$ (progenitor-based). The difference may be related to methodologies (excursion set formalism against fully coupled RHD here) or earlier reionization histories driven by brighter sources (global 0.5 reionization redshifts $z_R\sim 11$ instead of 7.8 here) yielding smaller lags between galaxies and the  IGM.

\subsection{Reionization Durations}

%

The spread $\Delta t$ of reionization times $t_\mathrm{reion}$ of particles in a halo at $z = 0$ can be used to compute the duration of its reionization. The result is plotted versus halo mass in Fig. \ref{fig:treion}. $\Delta t_{2\sigma}$ is computed from the r.m.s. of particle reionization times within a halo, using $\Delta t_{2\sigma}=(\langle t_\mathrm{part}\rangle+\sigma)-(\langle t_\mathrm{part}\rangle-\sigma)$. 

$\Delta t_{2\sigma}$ increases with halo mass, with typical values of $\sim 120$ Myr for $M_{z=0}>10^{11} h^{-1} M_\odot$. For $10^{12} h^{-1} M_\odot$, reionization durations as long as 180 Myr or as short as 60 Myr can be found. \citet{OCV14} made similar determinations for subhaloes of M31-MW analogs, and our results are consistent with their SPH model with similar emissivity for sources: our 120 Myr duration is typical of their reionization in isolated models, where inside-out reionization proceeds from inner regions of a galaxy to its outskirts.  Our shortest durations, $\Delta t_{2\sigma}=60$ Myr, are, on the other hand, typical of their externally-reionized scenario, where a nearby bright source 'flashes' the object. The scatter here reflects diverse environmental properties.

For $M_{z=0}>10^{11}h^{-1} M_\odot$, typical values of $\Delta t_{2\sigma}$ are comparable to the halo-to-halo scatter of $t_\mathrm{reion}$ and are likely to be lower bounds, since self-shielding may have been underestimated at our resolution limit. This is consistent with what \citet{ALV9} and \citet{LI14} found for more massive objects. Reionization was experienced at different times for different mass elements within any given present-day galaxy. This must be taken into account by any model of the impact of reionization on stellar populations. 

At lower mass, $M_{z=0}<10^{10}h^{-1} M_\odot$, objects have $\Delta t_{2\sigma}\sim 0$. This corresponds to extreme cases of fast external reionization or objects small enough to fit within a single cell of the reionization map (from AMR data smoothed to unrefined resolution $30 h^{-1}$ comoving kpc, comparable to the virial radius of a $\sim 10^9 h^{-1} M_\odot$ halo). Scatter increases toward the low-mass end, but these objects were sampled with a small number of cells or particles (40 particles for $10^8 h^{-1} M_\odot$). Their environmental history is not as fully resolved, leading to greater errors estimating $\Delta t_{2\sigma}$.

$\Delta t_{2\sigma}$ estimates reionization durations from the spread of times for typical particles, but the full range of time differences within a halo can be even greater.
The time difference $\Delta t_\mathrm{max-min}$ between the first and the last particle to reionize is also plotted in Fig. \ref{fig:treion}. The most massive halos can have $\Delta t_\mathrm{max-min} = 600$ Myr and thus contain material from locations that reionized at very different epochs.

Lower-mass haloes have smaller $\Delta t_\mathrm{max-min}$, with $\Delta t_\mathrm{max-min}\sim 400$ Myr for $10^{12} h^{-1}M_\odot$ but $\Delta t_\mathrm{max-min}< 100$ Myr for $M<10^{10} h^{-1} M_\odot$. For the lowest-mass galaxies, significant outliers are present, with $\Delta t_\mathrm{max-min}\sim 200$ Myr when the median value is closer to 20 Myr.


\subsection{The Local Group}

\begin{figure}[ht]
\includegraphics[width=1.2\columnwidth]{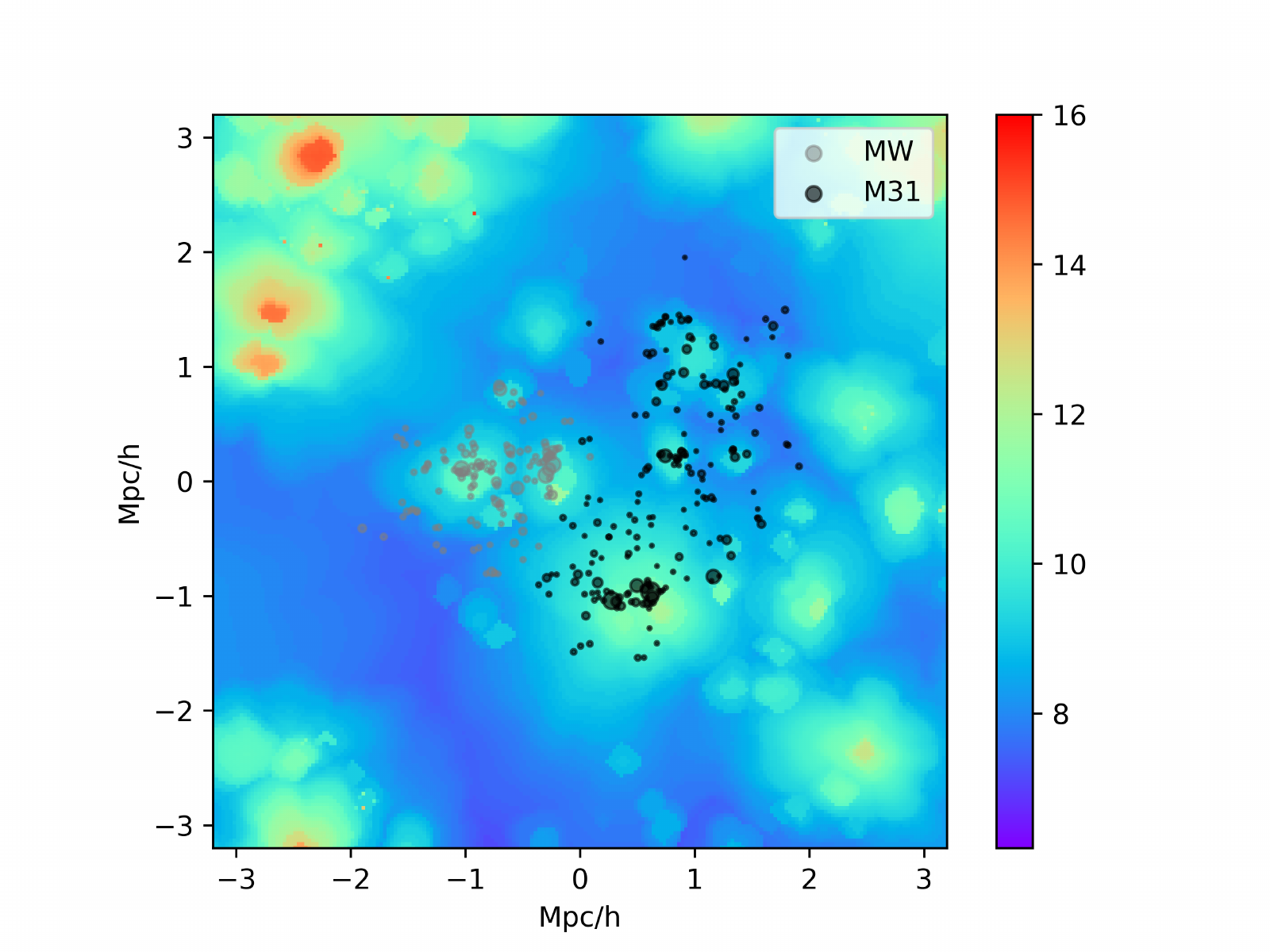}
\caption{A {5.7 Mpc z-axis projection} thru the CoDa I-AMR reionization map around the LG (background field), with the location of CoDa I-DM2048 halo progenitors at z=10.8 for simulated M31 and MW (symbols). Symbol sizes are proportional to halo masses. Background colors indicate {the maximal} reionization redshift {along the projection}.}
\label{fig:LG}
\end{figure}

Our CLUES ICs were constructed to form a LG, with two galaxies at $z = 0$ similar to the MW ($M_{z=0}=7.7\times 10^{11} h^{-1} M_\odot$) and M31 ($M_{z=0}=1.7 \times 10^{12} h^{-1} M_\odot$) at their correct separation, surrounded by a realistic large-scale environment that matches observations of the local universe. Fig. \ref{fig:LG} shows the positions of their progenitors at z=10.8 within the reionization map. The MW environment reionized in a compact fashion. The M31 reionization pattern, on the other hand, consists of several disconnected islands, reflecting the complex and extended distribution of progenitors at these times. Both objects reionized in isolation relative to each other: their patches are easily identified and disconnected. \cite{OCV11} and \cite{GIL15} predicted that this kind of reionization should lead to a more extended radial distribution of their satellites at z=0 compared to models without radiative transfer. The LG also reionized in isolation from the large-scale environment, as no ionization fronts appeared to sweep across them from outside. 

In Fig. \ref{fig:treion}, their reionization times and durations are shown for each estimator. By the progenitor-based method, the M31 environment reionized earlier($z=11$) than for MW ($z=9.8$), consistent with the general trend whereby haloes more massive started reionization earlier. By the particle-based method both objects reionized at the same time, at $z=8.2$~: since these two objects are spatially close, it is not surprising that their Lagrangian environments reionized simultaneously.  For both estimators, these two galaxies fall within the 5 - 95$\%$ contours, but they reionized later than the median for their masses. {Their measured mass growth is typical of halos of similar mass, suggesting that this delay is, instead, an environmental effect.}

Regarding durations, $\Delta t_{2\sigma}$ are typical of the global distribution, between $100$ and $150$ Myr. $\Delta t_{\mathrm max-min} \sim 400$ Myr are similar for the two objects, indicating their environments share similar extreme values for reionization times, presumably because of their proximity. 

These particle-based durations and times are consistent with the partially-suppressed model of \citet{DIX17}, where low-mass galaxies $M<10^9 h^{-1} M_\odot$ made a modest but 
non-negligible contribution to reionization~: it suggests a similar quantitative role for such objects in CoDa I-AMR.

 \section{Summary}
 
By combining a new, fully-coupled RHD simulation of reionization at $z > 6$ with an N-body simulation to $z = 0$ from the same ICs, we demonstrate that the redshifts at which present-day galaxies experienced reionization were correlated with their mass, by tracing their building-blocks back to the EOR. For $M_{z=0}$ between $10^{8}$ and $10^{13} h^{-1} M_\odot$, galaxies more massive than the MW were typically reionized earlier than the global mean, with a spread in reionization times for the building-blocks of a galaxy as large as galaxy-to-galaxy variations. This inhomogeneous timing of reionization amongst and within galaxies should be taken into account when modelling and interpreting stellar populations.  With CLUEs ICs, we modelled both global and LG reionization finding MW and M31 reionized earlier than the global mean {but later than galaxies of similar masses}, and without influence from outside the LG or each other.

\acknowledgments
DA, ND and PO acknowledge support from the French ANR funded project ORAGE (ANR-14-CE33-0016). PRS acknowledges the grant support of U.S. NSF AST-1009799, NASA NNX11AE09G, and DOE INCITE 2016 Award AST031 on the Titan supercomputer at Oak Ridge National Laboratory. ITI is supported by the UK Science and Technology Facilities Council [grant numbers ST/F002858/1 and ST/I000976/1] and The Southeast Physics Network (SEPNet).The CoDA I-DM2048 simulation was performed at LRZ Munich. GY would like to thank MINECO/FEDER (Spain) for financial support under research grant AYA2015-63810-P.

\end{document}